\documentclass[a4paper]{amsart}%
\usepackage{amssymb}
\usepackage{amsfonts}
\usepackage{graphicx}
\usepackage{breakcites}%
\usepackage{amsmath}%
\theoremstyle{plain}

\newtheorem{definition}{Definition}
\newtheorem{example}{Example}

\newtheorem{remark}{Remark}

\numberwithin{equation}{section}
\begin{document}
\title[On a composition of digraphs]{On a composition of digraphs}
\author{Simone Severini }
\address{Department of Computer Science, University of Bristol, BS8 1UB, Bristol,
United Kingdom}
\email{severini@cs.bris.ac.uk}
\date{March, 2003}

\begin{abstract}
Many \textquotedblleft good\textquotedblright\ topologies for interconnection
networks are based on line digraphs of regular digraphs. These digraphs
support unitary matrices. We propose the property \textquotedblleft being the
digraph of a unitary matrix\textquotedblright\ as additional criterion for the
design of new interconnection networks. We define a composition of digraphs,
which we call diagonal union. Diagonal union can be used to construct digraphs
of unitary matrices. We remark that diagonal union digraphs are state split
graphs, as defined in symbolic dynamics. Finally, we list some potential
directions for future research.

\end{abstract}
\maketitle

\section{Introduction}

Point-to-point interconnection networks for parallel and distributed systems
are usually modeled by (di)graphs. Survey papers on interconnection networks
are, \emph{e.g.}, (in chronological order)\ Feng \cite{F81}, Heydemann
\cite{H97} and Ferrero \cite{Fe99}. The vertices of the (di)graph correspond
to the nodes of the network, that is processors, nodes with local memory,
switches, \emph{etc}.; the arcs correspond to communication wires. Performance
of parallel and distributed systems is significantly determined by the choice
of the network topology. The basic requirements are: (i) large numbers of
nodes; (ii)\ small distance between any two nodes (small communication delay);
(iii) limited number of wires. These requirements result in the need of
finding (di)graphs, with, respectively: (i) large order; (ii) small diameter;
(iii) bounded degree. Another important requirement is fault-tolerance, which
is measured in the terms of several parameters, two of which are
vertex-connectivity and arc(edge)-connectivity.

Many \textquotedblleft good\textquotedblright\ topologies for interconnection
networks are based on line digraphs of regular digraphs (\emph{e.g.} de Bruijn
digraphs, Reddy-Pradhan-Kuhl digraphs, butterflies digraphs, \emph{etc.}),
hypercubes (that are not line digraphs), and a number of their
generalizations. Line digraphs have many remarkable properties. Line digraphs
are used as tools in algorithmic applications (see, \emph{e.g.}, Gusfield
\cite{GKWS98}), as underlying digraphs of coined quantum random walks
\cite{S03} (see Kempe \cite{K03}, for a survey paper on quantum random walks),
in the study of spectral statistics and random matrix theory (see,
\emph{e.g.}, Pakonski \emph{et al.} \cite{PTZ02}), in the study of growth
functions of free groups (see, \emph{e.g.}, Rivin \cite{R99}).

Recall that, given an $n\times n$ matrix $M$ (over any field), a digraph $D$
is said to \emph{support} $M$, or, equivalently, to be \emph{the digraph of
}$M$, if $D$ is on $n$ vertices, and its adjacency matrix, $M\left(  D\right)
$, has $ij$-th element
\[
M\left(  D\right)  _{i,j}=\left\{
\begin{tabular}
[c]{cc}%
$1$ & if $M_{i,j}\neq0,$\\
$0$ & otherwise.
\end{tabular}
\right.
\]
Recall that a complex $n\times n$ matrix $U$ is \emph{unitary} if $U^{\dagger
}U=I_{n}$, where $U^{\dagger}$ is the adjoint of $U$, and $I_{n}$ is the
identity matrix of size $n$. The digraph of a unitary matrix is without
cut-vertices and it is bridgeless, then it is $2$-vertex-connected and
$2$-arc-connected. Note that this properties insure a certain degree of
fault-tolerance when thinking about the digraph as a model of an
interconnection network. Both, line digraphs of regular digraphs and
hypercubes support unitary matrices \cite{S03I}. This observation motivates
the present paper, which is structured as follows.

A \emph{composition of digraphs} is a method to construct digraphs
\textquotedblleft putting together\textquotedblright\ smaller digraphs and
adding arcs according to specific rules. In \S 2 we define a composition of
digraphs, which we call \emph{diagonal union}. The composition can be used to
construct digraphs of unitary matrices. We observe: simple conditions under
which a diagonal union digraph supports unitary matrices; that line digraphs
of regular digraphs are diagonal union digraphs (\S 3); that diagonal union
digraphs are state split graphs, as defined in symbolic dynamics (\S 3).
Finally, in \S 4, we propose a few potential direction for future research.

Although the content of the paper is elementary, we wish to write it as
self-contained as possible. For terms of graph theory not defined here, we
refer the reader to the monograph by Bang-Jensen and Gutin \cite{B-JG01}.

\section{A composition of digraphs}

\subsection{Set-up}

A (finite) \emph{directed graph}, for short \emph{digraph}, consists of a
non-empty finite set of elements called \emph{vertices} and a (possibly
empty)\ finite set of ordered pairs of vertices called \emph{arcs}. Let us
denote by $D=\left(  V,A\right)  $ a digraph with vertex-set $V\left(
D\right)  $ and arc-set $A\left(  D\right)  $. The \emph{adjacency matrix} of
a digraph $D$ on $n$ vertices, denoted by $M\left(  D\right)  $, is the
$n\times n$ $\left(  0,1\right)  $-matrix with $ij$-th entry
\[
M_{i,j}\left(  D\right)  =\left\{
\begin{tabular}
[c]{ll}%
$1$ & if $\left(  v_{i},v_{j}\right)  \in A\left(  D\right)  ,$\\
$0$ & otherwise.
\end{tabular}
\ \right.
\]

A digraph $H$ is a \emph{subdigraph} of a digraph $D$ if $V\left(  H\right)
\subseteq V\left(  D\right)  $ and $A\left(  H\right)  \subseteq A\left(
D\right)  $. A subdigraph $H$ of a digraph $D$ is a \emph{spanning subdigraph}
of $D$ if $V\left(  H\right)  =V\left(  D\right)  $;\ in such a case, we say
that $H$ \emph{spans} $V\left(  D\right)  $. A \emph{decomposition} of a
digraph $D$ is a set $\left\{  H_{1},H_{2},...,H_{k}\right\}  $ of subdigraphs
whose arc-sets are exactly the classes of a partition of $A\left(  D\right)
$. A subdigraph of $D$ is called a \emph{factor} of $D$ if it spans $V\left(
D\right)  $. A \emph{factorization} of $D$ is a decomposition of $D$ into factors.

The \emph{Hadamard product} (in the literature, sometimes called \emph{Schur
product} or \emph{entry-wise product}) of $n\times m$ matrices $N$ and $M$,
denoted by $N\circ M$, is defined as follows:
\[
\left(  N\circ M\right)  _{i,j}=N_{i,j}\cdot M_{i,j}.
\]

\begin{definition}
[Generalized Hadamard product]Let $M$ and $N$ be respectively an $m\times m$
and an $n\times n$ matrix (over any field). Let $m=n\cdot r$. The
\emph{generalized Hadamard product} of $N$ with $M$, denoted by $N\circ_{G}M$,
is defined as follows:

\begin{itemize}
\item[(i)] If $r=1$ ($n=m$ and $N$ has the same size of $M$) then
\[
N\circ_{G}M=N\circ M.
\]

\item[(ii)] If $r>1$ then we look at $M$ as a block-matrix of the form
\[
\left[
\begin{array}
[c]{ccc}%
A_{1,1} & \cdots & A_{1,n}\\
\vdots & \ddots & \vdots\\
A_{n,1} & \cdots & A_{n,n}%
\end{array}
\right]  ,
\]
where the $ij$-th block is $r\times r$. Then
\[
N\circ_{G}\left[
\begin{array}
[c]{ccc}%
A_{1,1} & \cdots & A_{1,n}\\
\vdots & \ddots & \vdots\\
A_{n,1} & \cdots & A_{n,n}%
\end{array}
\right]  =\left[
\begin{array}
[c]{ccc}%
N_{1,1}A_{1,1} & \cdots & N_{1,n}A_{1,n}\\
\vdots & \ddots & \vdots\\
N_{n,1}A_{n,1} & \cdots & N_{n,n}A_{n,n}%
\end{array}
\right]  .
\]

\end{itemize}
\end{definition}

\begin{example}
Let
\[
\sigma_{x}=\left[
\begin{array}
[c]{cc}%
0 & 1\\
1 & 0
\end{array}
\right]  .
\]
Then
\[%
\begin{tabular}
[c]{lll}%
$\sigma_{x}\circ_{G}\left[
\begin{array}
[c]{cc}%
A & B\\
D & C
\end{array}
\right]  =\left[
\begin{array}
[c]{cc}%
\mathbf{0} & B\\
D & \mathbf{0}%
\end{array}
\right]  $ & and & $I_{2}\circ_{G}\left[
\begin{array}
[c]{cc}%
A & B\\
D & C
\end{array}
\right]  =\left[
\begin{array}
[c]{cc}%
A & \mathbf{0}\\
\mathbf{0} & C
\end{array}
\right]  .$%
\end{tabular}
\]

\end{example}

\subsection{Definition}

Let $G$ be a finite group of order $n$. Let $W$ be a complex vector space of
finite dimension. Let $GL\left(  W\right)  $ be the group of the bijective
linear maps on $V$. A \emph{linear representation} of $G$ is a homomorphism
$\rho:G\longrightarrow GL\left(  W\right)  $. To each $g_{i}\in G$, we
associate the unit column vector $e_{i}\in W$ with $j$-th entry
\[
e_{i_{j}}=\left\{
\begin{tabular}
[c]{ll}%
1 & if $j=i,$\\
0 & otherwise.
\end{tabular}
\ \right.
\]
Let $GL\left(  n,F\right)  $ be the group of invertible $n\times n$ matrices
over the field $F=GF\left(  2\right)  $. The (left) \emph{regular permutation
representation} of $G$, denoted by $\rho_{reg}$, is a homomorphism $\rho
_{reg}:G\longrightarrow GL\left(  n,F\right)  $. Then $\rho_{reg}\left(
g_{l}\right)  $ is a permutation matrix with $ij$-entry
\[
\rho_{reg}\left(  g_{l}\right)  _{i,j}=\left\{
\begin{tabular}
[c]{ll}%
1 & if $g_{l}g_{i}=g_{j},$\\
0 & otherwise.
\end{tabular}
\ \right.
\]

Let us denote by $\mathbb{Z}_{n}$ the additive abelian group of the integers
modulo $n$. Let us denote by $J_{n}$ the $n\times n$ all-ones matrix, and by
$M\oplus N$ the direct sum of matrices $M$ and $N$.

\begin{definition}
[Diagonal union]Let $D$ be a digraph on $n$ vertices and let $\mathcal{F}%
=\left\{  H_{1},H_{2},...,H_{k}\right\}  $ be a factorization of $D$. The
digraph $D_{\mathcal{F}}$ is defined as follows:
\[
M\left(  D_{\mathcal{F},1}\right)  =M\left(  D_{\mathcal{F}}\right)  =\left(
J_{k}\otimes I_{n}\right)  \cdot\bigoplus_{i=1}^{k}M\left(  H_{i}\right)  .
\]
Iteratively,
\[
M\left(  D_{\mathcal{F},d}\right)  =\left(  J_{k}\otimes I_{n^{d-1}}\right)
\cdot\bigoplus_{j=0}^{k-1}\left[  \rho_{reg}\left(  j\right)  \circ
_{G}M\left(  D_{\mathcal{F},d-1}\right)  \right]  ,
\]
$\rho_{reg}$ is the (left)\ regular permutation representation of
$\mathbb{Z}_{k}$. The digraph $D_{\mathcal{F},d}$ is called $d$\emph{-diagonal
union}\footnote{On the choice of the terminology:\ The term \textquotedblleft
diagonal union\textquotedblright\ may be justified by the structure of
$M\left(  D_{\mathcal{F},d}\right)  $. In fact, $M\left(  D_{\mathcal{F}%
,d}\right)  $ is a block-matrix, with the blocks on the diagonal being the
adjacency matrices of the factors of $D_{\mathcal{F},d-1}$.} of $D$ in respect
to $\mathcal{F}$.
\end{definition}

\begin{remark}
If the factors $H_{1},H_{2},...,H_{k}$ are digraphs of unitary matrices then
$M\left(  D_{\mathcal{F},d}\right)  $ is also the digraph of a unitary matrix.
This insures that $M\left(  D_{\mathcal{F},d}\right)  $ is $2$%
-vertex-connected and $2$-arc-connected \cite{S03I}.
\end{remark}

A concrete example might be useful to clarify Definition 2. Let us denote by
$\overleftrightarrow{K}_{n}^{+}$ the complete symmetric digraph with a
self-loop at each vertex.

\begin{example}
Let $\mathcal{F}=\left\{  H_{1},H_{2}\right\}  $ be a factorization of a
digraph $D$. In Figure \ref{fig1}, below, are represented the matrices
$M\left(  D_{\mathcal{F}}\right)  ^{\intercal}$, $M\left(  D_{\mathcal{F}%
,2}\right)  ^{\intercal}$, $M\left(  D_{\mathcal{F},3}\right)  ^{\intercal}$
and $M\left(  D_{\mathcal{F},4}\right)  ^{\intercal}$. The light and the dark
shaded blocks represent $M\left(  H_{1}\right)  $ and $M\left(  H_{2}\right)
$, respectively; the white areas are of zeros only.

Let
\[%
\begin{tabular}
[c]{llll}%
$D=\overleftrightarrow{K}_{2}^{+},$ & $M\left(  H_{1}\right)  =I_{2}$ & and &
$M\left(  H_{2}\right)  =\sigma_{x}.$%
\end{tabular}
\]
Then
\[
M\left(  D_{\mathcal{F},k}\right)  =B\left(  2,k\right)  ,
\]
the binary $k$-dimensional de Bruijn digraph.%
\begin{figure}
[ptb]
\begin{center}
\includegraphics[
natheight=17.078800cm,
natwidth=11.508100cm,
height=6.0297cm,
width=4.0857cm
]%
{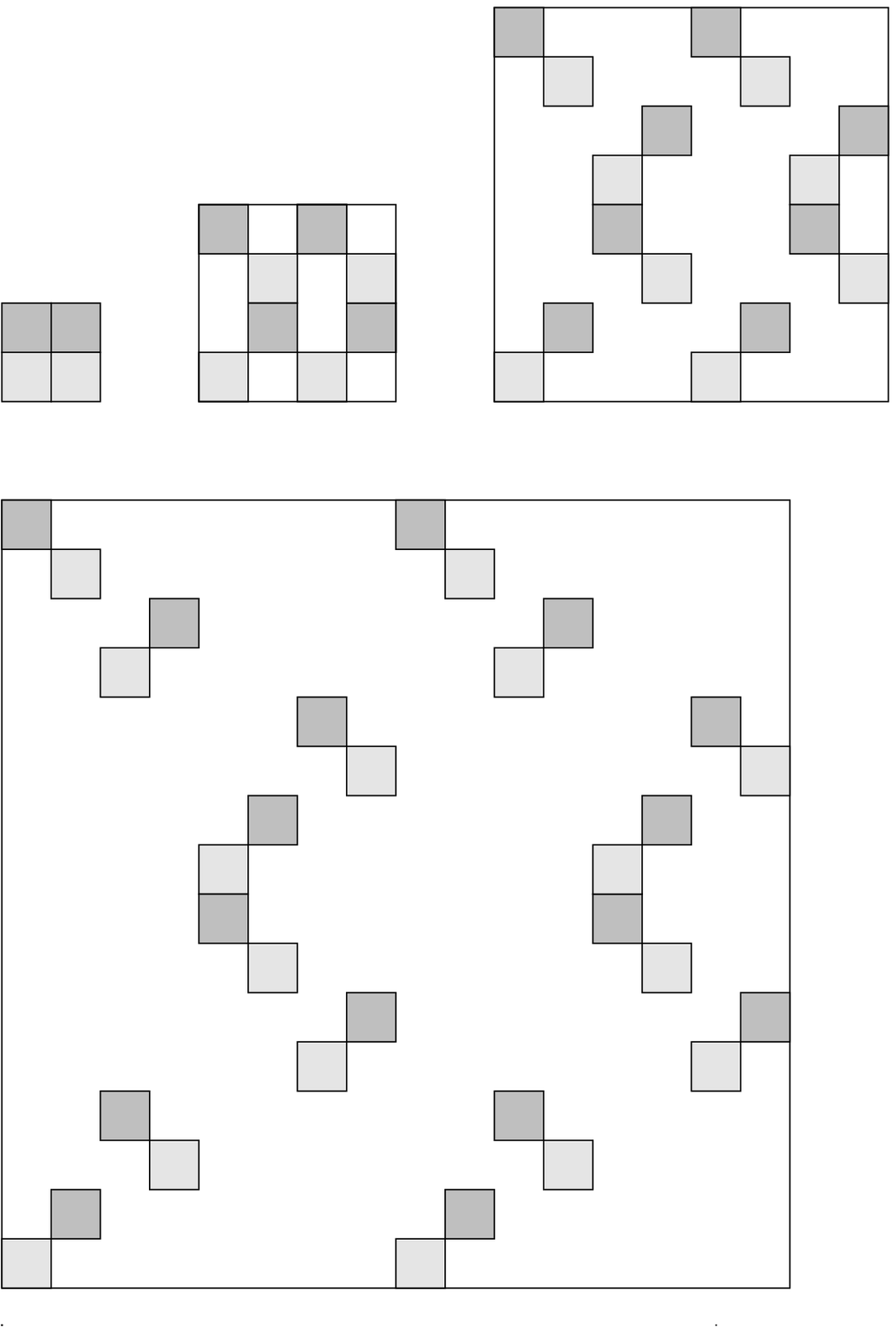}%
\label{fig1}%
\end{center}
\end{figure}

\end{example}

\section{Remarks}

\subsection{Extremal cases}

Let $D$ be a digraph on $n$ vertices and $m$ arcs. Suppose that $\mathcal{F}$
is the trivial factorization of $D$, that is $\mathcal{F}=\left\{  D\right\}
$. In such a case, $D_{\mathcal{F}}=D$. A \emph{cycle factor} of a digraph $D$
is a collection of pairwise vertex-disjoint dicycles spanning $D$. In other
words, a cycle factor is a spanning $1$-regular subdigraph of $D$. Note that
the adjacency matrix of a cycle factor is a permutation matrix. A
\emph{factorization in cycle factors} is a factorization whose members are
cycle factors. The \emph{line digraph} of a digraph $D$, denoted by
$\overrightarrow{L}D$, is defined as follows (see, \emph{e.g.}, \cite{P96}%
):\ the vertex-set of $\overrightarrow{L}D$ is $A\left(  D\right)  $; $\left(
v_{i},v_{j}\right)  ,\left(  v_{k},v_{l}\right)  \in A\left(  \overrightarrow
{L}D\right)  $ if and only if $v_{j}=v_{k}$. Suppose that $\mathcal{F}$ is a
factorization in cycle factors. In such a case,
\[
D_{\mathcal{F},d}=\overrightarrow{L}^{d-1}D.
\]
So, note that the line digraph of a regular digraph is a diagonal union digraph.

\subsection{State split graphs}

The notion of \emph{state split graph}, introduced by Williams \cite{W74}, is
important in symbolic dynamics and coding. Lind and Marcus \cite{LM95} is
comprehensive for a monograph on symbolic dynamics. Given a digraph $D$, let
$\mathcal{E}_{i}$ be the set of the arcs in which $v_{i}$ is tail. For each
vertex $v_{i}\in V\left(  D\right)  $, let
\[%
\begin{tabular}
[c]{lll}%
$\mathcal{E}_{i}=\biguplus_{k=1}^{m\left(  i\right)  }\mathcal{E}_{i}^{k},$ &
& $m\left(  i\right)  \geq1.$%
\end{tabular}
\]
Let $\mathcal{P}$ be the resulting partition of $\mathcal{E}$, and let
$\mathcal{P}_{i}$ the resulting partition restricted to $\mathcal{E}_{i}$. The
\emph{state split graph} $D^{\left[  \mathcal{P}\right]  }$ \emph{formed from}
$D$ \emph{using} $\mathcal{P}$ has vertices $v_{i}^{1},v_{i}^{2}%
,...,v_{i}^{m\left(  i\right)  }$, for all $v_{i}\in V\left(  D\right)  $. The
arc $\left(  v_{h},v_{l}\right)  ^{j}\in A\left(  D^{\left[  \mathcal{P}%
\right]  }\right)  $, where $\left(  v_{h},v_{l}\right)  \in A\left(
D\right)  $ and $1\leq j\leq m\left(  v_{l}\right)  $. The arc $\left(
v_{h},v_{l}\right)  \in\mathcal{E}_{h}^{k}$, for some $k$. The tail and the
head of $\left(  v_{h},v_{l}\right)  ^{j}$ are respectively $v_{h}^{k}$ and
$v_{l}^{j}$. If the construction of a state split graph uses a partition of
the outgoing (ingoing) arcs (as above) then the obtained digraph is called
\emph{out-split (in-split) graph}.

\begin{remark}
A digraph $M\left(  D_{\mathcal{F}}\right)  $ is a state split graph. In
particular, $M\left(  D_{\mathcal{F}}\right)  $ is a in-split graph formed
from $D$ using the partition of $A\left(  D\right)  $ arising from the
factorization $\mathcal{F}$.
\end{remark}

\begin{example}
Consider the Cayley digraph $D=Cay\left(  \mathbb{Z}_{4},\left\{
1,2,3\right\}  \right)  $:
\[
M\left(  D\right)  =\sum_{i=1}^{3}\rho_{reg}\left(  i\right)  =\left[
\begin{array}
[c]{ccccc}%
\emph{0} & 0 & 1 & 1 & 1\\
\emph{1} & 1 & 0 & 1 & 1\\
\emph{2} & 1 & 1 & 0 & 1\\
\emph{3} & 1 & 1 & 1 & 0
\end{array}
\right]  ,
\]
where in the first column are the labels of the vertices.

Let $\mathcal{F}=\left\{  D_{1},D_{2}\right\}  $, where
\[%
\begin{tabular}
[c]{lll}%
$M\left(  D_{1}\right)  =\rho_{reg}\left(  1\right)  +\rho_{reg}\left(
3\right)  $ & and & $M\left(  D_{2}\right)  =\rho_{reg}\left(  2\right)  .$%
\end{tabular}
\]
Then
\begin{align*}
M\left(  D_{\mathcal{F}}\right)   & =\left[  \rho_{reg}\left(  1\right)
\otimes\rho_{reg}\left(  3\right)  \right]  \cdot\left(  J_{2}\otimes
I_{4}\right) \\
& =\left[
\begin{array}
[c]{ccccccccc}%
\emph{0}^{\emph{1}} & 0 & 1 & 0 & 1 & 0 & 0 & 1 & 0\\
\emph{1}^{\emph{1}} & 1 & 0 & 1 & 0 & 0 & 0 & 0 & 1\\
\emph{2}^{\emph{1}} & 0 & 1 & 0 & 1 & 1 & 0 & 0 & 0\\
\emph{3}^{\emph{1}} & 1 & 0 & 1 & 0 & 0 & 1 & 0 & 0\\
\emph{0}^{\emph{2}} & 0 & 1 & 0 & 1 & 0 & 0 & 1 & 0\\
\emph{1}^{\emph{2}} & 1 & 0 & 1 & 0 & 0 & 0 & 0 & 1\\
\emph{2}^{\emph{2}} & 0 & 1 & 0 & 1 & 1 & 0 & 0 & 0\\
\emph{3}^{\emph{2}} & 1 & 0 & 1 & 0 & 0 & 1 & 0 & 0
\end{array}
\right]  .
\end{align*}
Then $D_{\mathcal{F}}$ $\ $is an in-split graph formed from $D$ using the
partition arising from $\mathcal{F}$.
\end{example}

\section{What's next?}

The following seem to be natural questions:

\begin{itemize}
\item[(i)] To what extent the property ``being the digraph of a unitary
matrix'' makes a digraph a ``good'' network topology?

\item[(ii)] Can we make use of diagonal union digraphs as bases of new network topologies?

\item[(iii)] We have seen that diagonal union digraphs are obtained via
state-splitting. Are tools from symbolic dynamics and coding helpful in the
design and analysis of interconnection networks, and in broadcasting problems?
\end{itemize}

There is also space for a legitimate speculation:

\begin{itemize}
\item[(i)] The time-evolution of a quantum mechanical system, in a pure state,
is induced by unitary matrices. Then interconnection networks based on
digraphs of unitary matrices have something to do with quantum mechanical
systems. What? In some way, these networks model certain classical systems,
whose dynamics share some combinatorial structure with quantum systems. So,
they are (partially?) combinatorially quantum. Can this idea be exploited in
the context of quantum computation and communication?
\end{itemize}

\bigskip

\textbf{Acknowledgement.} The author wishes to thank Yaokun Wu for introducing
him to symbolic dynamics, Cecilia Bebeacua and Dhiraj Pradhan for showing
interest in this paper. The author is supported by a University of Bristol
Research Scholarship.

\end{document}